\documentclass[aps,prx,twocolumn,superscriptaddress,floatfix]{revtex4-1}
\usepackage[utf8]{inputenc}
\usepackage{graphicx}
\usepackage{amsmath}
\usepackage{amssymb}
\usepackage{braket}
\usepackage{cancel}
\usepackage{algorithm,algpseudocode}

\usepackage[colorlinks=true, citecolor={blue!80!black}, urlcolor={blue!50!black}, linkcolor = {blue!80!black}]{hyperref}

\usepackage{xcolor}

\newcommand{\norm}[1]{\left\lVert#1\right\rVert}

\newcommand{\gfmps}{g\! f\! m\hspace{-0.07em}ps}

\AtBeginDocument{
    \DeclareMathOperator{\Tr}{Tr}
    \DeclareMathOperator{\Pf}{Pf}
}

\begin{document}

\title{Linear-time generalized Hartree-Fock algorithm for quasi-one-dimensional systems}

\author{Alex Meiburg}
\affiliation{Department of Physics, University of California, Santa Barbara, CA 93106, USA}

\author{Bela Bauer}
\affiliation{Microsoft Station Q, Santa Barbara, California 93106 USA}

\begin{abstract}

    In many approximate approaches to fermionic quantum many-body systems, such as Hartree-Fock and density functional theory, solving a system of non-interacting fermions coupled to some effective potential is the computational bottleneck. In this paper, we demonstrate that this crucial computational step can be accelerated using recently developed methods for Gaussian fermionic matrix product states (GFMPS). As an example, we study the generalized Hartree-Fock method, which unifies Hartree-Fock and self-consistent BCS theory, applied to Hubbard models with an inhomogeneous potential. We demonstrate that for quasi-one-dimensional systems with local interactions, our approach scales approximately linearly in the length of the system while yielding a similar accuracy to standard approaches that scale cubically in the system size.
\end{abstract}

\maketitle

\section{Introduction}

While the quantum many-body problem for fermions can in general not be solved numerically efficiently, a plethora of approximate computational approaches have been developed that are able to capture relevant properties of many-fermion systems in certain limits. A widely used class of such methods are Hartree-Fock and self-consistent Bardeen-Cooper-Shrieffer (BCS) theory~\cite{bardeen1957microscopic,bardeen1957theory}. These can be viewed as variational mean-field methods: they find the best approximation to the exact ground state within the space of non-interacting fermion states. Another powerful approach is density-functional theory, which expresses the total energy of the fermionic system as a -- generally unknown -- functional of the density~\cite{hohenberg1964}; while finding this functional is difficult, decades of numerical experience have shown that relatively simple approximations to this functional, such as the local density approximation, can successfully describe materials where the effect of interactions is moderate~\cite{jones2015}.

Common to these approaches is that the most computationally expensive step in the numerical simulation is finding the ground state of a system of fermions coupled to an effective potential (the mean-field potential in the case of HF and BCS~\cite{hartree1935self}, and the Kohn-Sham potential in the case of DFT~\cite{kohn1965self}). Without further approximation, this step scales cubically in the number of degrees of freedom and as such becomes prohibitively costly for systems in excess of a few thousand degrees of freedom.

Here, we will demonstrate that in low spatial dimensions and for local Hamiltonians, this step can be accelerated significantly by using tensor network states (TNS)~\cite{white1992,white1992-1,fannes1992,ostlund1995,sierra1998,nishino1998,nishino2000,nishio2004,verstraete2004,jordan2008,vidal2007-1,vidal2008} (for recent reviews, see Refs.~\cite{orus2014,orus2014-1,bridgeman2017}). Such states are known to be able to compactly represent weakly entangled quantum many-body states, such as the ground states of local Hamiltonians. In many cases, the computational scaling of these approaches is empirically found to be approximately linear in the size of the system and exponential in its bipartite entanglement. However, in the case of free fermions, this can be reduced further to a polynomial scaling in the entanglement by using so-called Gaussian fermionic tensor network states~\cite{kraus2010,evenbly2010entanglement,schuch2012,dubail2015,haegeman2013,fishman2015,evenbly2016,haegeman2018,jahn2019holography}. Recently, a particular variant, Gaussian fermionic matrix-product states (GFMPS)~\cite{schuch2012}, was used as basis for efficient computational methods for non-interacting fermions in quasi-one-dimensional systems. These methods are able to compute equilibrium and non-equilibrium properties for systems order of magnitudes larger than naive approaches~\cite{schuch2019}.

In this paper, we focus on accelerating the self-consistent generalized Hartree-Fock (gHF) iteration~\cite{bach1994generalized,bogo58,Dichtel1971,gennesBook99} using Gaussian fermionic matrix-product states. We begin by reviewing the gHF approach, which should be viewed as the most general variational method using states of non-interacting fermions and elegantly unifies Hartree-Fock and self-consistent BCS theory. We then rederive the self-consistency equations for gHF, review key properties of GFMPS and discuss how to efficiently implement the gHF iteration using GFMPS. Finally, we demonstrate the approach on an example of interacting fermions in a quasi-dimensional geometry in the presence of an inhomogeneous trapping potential.

We note that other numerical approaches to solve the gHF problem have been discussed in Ref.~\onlinecite{kraus2010-ghf,kraus2013gHF}. In particular, Ref.~\onlinecite{kraus2010-ghf} discusses how to perform real- and imaginary-time evolution in the gHF setting, and pursues imaginary-time evolution as an approach to find the ground states. We focus here instead on the self-consistent field approach, which is often faster but may be more prone to becoming trapped in local minima. We note that the time evolution described in Ref.~\onlinecite{kraus2010-ghf} could similarly be accelerated using GFMPS techniques~\cite{schuch2019}. A different approach to improve the performance of gHF based on highly scalable methods for solving the fermionic problem for sparse systems was discussed in Ref.~\onlinecite{LinWu2021}, reaching remarkably large systems by parallelizing the computation on several thousand computational cores.

\section{Methods}

\subsection{Gaussian fermionic states}

We consider a lattice of fermions, where the operators $a_i^\dagger$ and $a_i$ create and annihilate a fermion on the $i$'th site, respectively.
For our purposes, it will be convenient to introduce a basis of self-adjoint Majorana fermion operators, which we denote $c_i$, and which satisfy the commutation relations $\{c_i,c_j\} = 2\delta_{ij}$. They are related to the standard creation and annihilation operators by $c_{2i-1} = a_i + a_i^\dagger$, $c_{2i}=-i(a_i - a_i^\dagger)$. In this way, any system with $N$ fermionic modes can be rewritten as one with $2N$ Majorana fermions.

Gaussian states are the states with density matrix $\rho$ proportional to $e^{-i {\bf c} A {\bf c}}\ket{0}$~\cite{bravyi2004}, where $\bf c$ is the vector of Majorana operator $c_i$, $A$ is a real anti-symmetric matrix, and $\ket{0}$ is the fermionic vacuum. Any fermionic state $\rho$ has an associated real anti-symmetric covariance matrix $\Gamma$ given by 
\begin{equation}
    \Gamma_{ij}=\frac{i}{2}\Tr(\rho[c_i,c_j]).
\end{equation}
For pure states, which we focus on here, $\Gamma$ has to satisfy $\Gamma^2 = -\openone$. For Gaussian states, this matrix contains a full description as the expectation value of any operator can be computed from it~\cite{Bravyi2017}: the expectation value of a Majorana monomial $\prod_{x \in \mathcal{X}} c_x$, where $\mathcal{X}$ denotes some set of lattice sites, is given by
\begin{equation} \label{eqn:expval}
    \Tr\left(\rho \prod_{x \in \mathcal{X}} c_x\right) = \Pf(\Gamma_\mathcal{X}),
\end{equation}
where $\Pf(\cdot)$ denotes the Pfaffian and $\Gamma_\mathcal{X}$ the covariance matrix restricted to the sites in $\mathcal{X}$. This should be viewed as embodiment of Wick's theorem in the covariance matrix formalism.

For Hamiltonians quadratic in creation and annihilation operators (alternately, quadratic in Majorana operators), i.e. of the form
\begin{equation}
    \mathcal{H} = -i \sum H_{ij} c_i c_j
\end{equation}
where $H$ is real and anti-symmetric, the ground state is always Gaussian.
$H$ can be diagonalized and has purely imaginary eigenvalues. The minimum energy state is determined by the covariance matrix
\begin{equation}
\Gamma = i(V_- V_-^\dagger - V_+ V_+^\dagger)
\end{equation}
where $V_+$ ($V_-$) are the normalized eigenvectors of $H$ corresponding to eigenvalues with positive (negative) imaginary parts. In this way, $\Gamma$ and the ground state can be determined through a diagonalization of $H$. Such diagonalization takes $\mathcal{O}(N^3)$ time.

Every pure Gaussian fermionic has well-defined fermion parity, i.e. any pure Gaussian state $\ket{\gamma}$ satisfies $\prod c_i \ket{\gamma} = e^{i \pi \sum a_i^\dagger a_i} \ket{\gamma} = p \ket{\gamma}$ with $p = \pm 1$. In terms of the covariance matrix $\gamma$ corresponding to this state, the parity is given by $\Pf(\gamma)$, where $\Pf$ denotes the Pfaffian.
An important subset of states within the class of Gaussian fermionic states are those with a well-defined particle number, i.e. that satisfy $\sum a_i^\dagger a_i \ket{\gamma} = n \ket{\gamma}$ for some integer $n$. These are traditionally referred to as Slater determinants and can be written as $\prod d_i^\dagger \ket{0}$, where the $d_i^\dagger$ are a new set of fermionic creation operators that are related to the original $a_i^\dagger$ by a unitary transformation, and $\ket{0}$ is again the fermionic vacuum.

\subsection{Generalized Hartree-Fock}

For Hamiltonians that are not quadratic, finding the ground state is in general exponentially difficult. However, the solution can be approximated using a variational approach, i.e. finding the state within some efficiently parametrized variational class that minimizes the expectation value of the energy. If a sufficiently powerful class of variational states is chosen, this approximates physical properties of the true ground state accurately.
Choosing this variational class to be the Slater determinants, i.e. Gaussian fermionic states with fixed particle number, leads to the well-known Hartree-Fock approach. By considering the entire set of Gaussian fermionic states, i.e. including those with fluctuating particle number, one arrives at a generalized Hartree-Fock approach that is also able to capture superconductivity at the mean-field level, i.e. contains the ground states of BCS theory where the superconducting order parameter has no quantum fluctuations~\cite{bach1994generalized}. It is known that there exist systems that are much better approximated by generalized Hartree-Fock than by non-generalized Hartree-Fock~\cite{BGKTApproximation2019}.


We now review this generalized Hartree-Fock approach, rederive the self-consistent iteration for its numerical solution, and clarify its relation to better-known approaches. Our starting point is a Hamiltonian that is quartic in the fermion operators:
\begin{equation}
    \mathcal{H} = -i\sum T_{ij} c_i c_j + \sum U_{ijk\ell} c_i c_j c_k c_l.
\end{equation}
Here, $T_{ij}$ is real and antisymmetric, while $U_{ijk\ell}$ is real and antisymmetric under exchange of any two indices, i.e. $U_{ijk\ell} = -U_{jik\ell} = U_{jki\ell} = \ldots$. Any quartic fermion Hamiltonian can be written in this form, including physically relevant cases such as the Hubbard Hamiltonian and the Coulomb interaction (see Appendix~\ref{appendixMaj} for details).

The energy for a Gaussian state $\ket{\Gamma}$ with corresponding covariance matrix $\Gamma$ is easily evaluated using Eqn.~\eqref{eqn:expval} by recognizing that the expectation value of the 4-fermion term is given by
\begin{eqnarray}
\braket{c_i c_j c_k c_l} &=& \Pf \begin{pmatrix}
    0 &\Gamma_{ij} &\Gamma_{ik} &\Gamma_{il} \\
    -\Gamma_{ij} &0 &\Gamma_{jk} &\Gamma_{jl} \\
    -\Gamma_{ik} &-\Gamma_{jk} &0 &\Gamma_{kl} \\
    -\Gamma_{il} &-\Gamma_{jl} &-\Gamma_{kl} &0
\end{pmatrix} \\
&=& \Gamma_{ij} \Gamma_{kl} - \Gamma_{ik} \Gamma_{jl} + \Gamma_{il} \Gamma_{jk}.
\end{eqnarray}
We can now use the identites
\begin{multline} \label{eqn:id1}
- \sum U_{ijkl} \Gamma_{ik} \Gamma_{jl} = - \sum U_{ikjl} \Gamma_{ij} \Gamma_{kl} \\ = \sum U_{ijkl} \Gamma_{ij} \Gamma_{kl}
\end{multline}
and
\begin{multline} \label{eqn:id2}
\sum U_{ijkl} \Gamma_{il} \Gamma_{jk} = \sum U_{iljk} \Gamma_{ij} \Gamma_{kl} \\ = \sum U_{ijkl} \Gamma_{ij} \Gamma_{kl}
\end{multline}
to arrive at the final expression
\begin{equation}
    \bra{\Gamma} \mathcal{H} \ket{\Gamma} = \sum T_{ij}\Gamma_{ij} + 3\sum U_{ijk\ell}\Gamma_{ij}\Gamma_{kl}.
\end{equation}
The factor of 3 can be viewed as counting the Hartree term, Fock term, and BCS term each, which are traditionally viewed as distinct. However, due to the symmetries of the Majorana representation, here these three terms all appear symmetrically.



We now have to find the pure-state covariance matrix $\Gamma$ (i.e. satisfying the non-linear constraint $\Gamma^2=-\openone$) that minimizes the above expression. We can recover the typical self-consistent HF iteration by starting with an initial guess $\Gamma^0$ (satisfying $(\Gamma^0)^2 = -\openone$) and expressing the new state as $\Gamma = \Gamma^0 + \delta \Gamma$ (where $\delta \Gamma$ is not by itself a valid covariance matrix). In terms of this new $\Gamma$ and the starting point $\Gamma^0$, the energy is given by
\begin{eqnarray}
     \braket{\mathcal{H}}
     & = & \sum T_{ij}\Gamma_{ij}^0 + 3\sum U_{ijk\ell}\Gamma_{ij}^0\Gamma_{k\ell}^0\\
     & & + \sum T_{ij}(\Gamma_{ij} - \Gamma_{ij}^0) + 3\sum U_{ijk\ell}(\Gamma_{ij} - \Gamma_{ij}^0)\Gamma^0_{k\ell}\notag \\
     & & + 3\sum U_{ijk\ell}\Gamma^0_{ij}(\Gamma_{k\ell} - \Gamma_{k\ell}^0) + O(\norm{\Gamma - \Gamma^0}^2)\notag \\
     & \approx & \mathrm{const.} + \sum \left[ T_{ij}+6U_{ijk\ell}\Gamma^0_{k\ell} \right] \Gamma_{ij}
\end{eqnarray}
Here, we have made the key approximation to neglect terms of order $\norm{\Gamma - \Gamma^0}^2$ in order to arrive at a linear functional of $\Gamma$. Furthermore, we have used the same symmetries as in Eqns.~\eqref{eqn:id1},~\eqref{eqn:id2} to collect different terms together. We note that the final expression can be viewed as an effective quadratic Hamiltonian acting on $\Gamma$,
\begin{equation}
    F_{ij} = T_{ij} +6\sum_{k\ell}U_{ijk\ell}\Gamma^0_{k\ell},
\end{equation}
which is commonly referred to as Fock matrix. Its ground state is by construction a valid covariance matrix that satisfies $\Gamma^2 = -\openone$. When the system has local hopping and interaction terms $T$ and $U$, they are sparse and will have only $O(N)$ entries, so the Fock matrix can be computed from $\Gamma$ in $O(N)$ time.
The iteration now proceeds by solving for the ground state of $F$, then replacing $\Gamma^0$ by that new state and recomputing the Fock matrix $F$, and repeating this procedure until convergence.

While it is known that this iteration cannot find the lowest-energy state in all cases~\cite{schuch2007computational}, for many systems, especially those in which $\left.\norm{U} \ll \norm{T}\right.$, it is empirically known to converge rapidly and reliably to a global minimum energy. In other cases, there can be local minima or stable oscillations. The Optimal Damping Algorithm attempts to remediate this by choosing the minimum-energy convex combination $t \Gamma^{\rm{new}} + (1-t)\Gamma^{\rm{old}}$~\cite{ODA}. Since the energy is a scalar quadratic function of $t$, this can be directly minimized through evaluation at any three values of $t$.

Computing $\Gamma$ from $H$ requires an eigenvalue decomposition, an operation which scales as $O(N^3)$ in general. In many cases, this will become the computational bottleneck and limit the system size for which Hartree-Fock can be used to several fermionic degrees of freedom. It is worth noting, however, that there are important use-cases for Hartree-Fock where this is not the bottleneck. For example, in quantum chemistry the Hamiltonian is non-local and the basis is not a real-space grid, such that there are $\mathcal{O}(N^4)$ terms in the Hamiltonian that need to be computed as multi-dimensional integrals over the basis functions. In this case, computing the terms of the Hamiltonian is the bottleneck of the Hartree-Fock simulation.
However, as we will see in the next section, when the Hamiltonian is local and the system quasi-one-dimensional, Gaussian fermionic tensor networks offer a more time- and memory-efficient approach.

\subsection{Gaussian fermionic tensor networks}

While a generic state on $N$ particles can have as many as $N/2$ bits of entanglement across a cut, obeying what is commonly referred to as a volume law, the entanglement in low-energy states is typically much less. The situation is best understood for gapped, local Hamiltonians in one spatial dimension, which are known to have area-law entanglement in their ground state~\cite{Eisert2010AreaLF}, i.e. the entanglement is bounded by a constant regardless of system size. The same behavior is expected for most systems also in higher dimensions~\cite{acoleyen2013entanglement}. The area law is typically violated in gapless systems; however, in many cases this violation is mild. For example, conformal field theories in 1D have only $O(\log(N))$ entanglement~\cite{holzhey1994,vidal2003}, i.e. the area law is violated by a logarithmic correction.

Tensor networks~\cite{orus2014,orus2014-1,bridgeman2017} make use of the entanglement properties of low-energy states to represent them more efficiently.
Matrix product states (MPS) are a particular class of tensor network states~\cite{fannes1992,white93,ostlund1995} that is known to be able to efficiently represent the ground states of gapped one-dimensional Hamiltonians. Furthermore, MPS can be manipulated efficiently and the variational problem can in many cases be solved efficiently using the density-matrix renormalization group (DMRG) method~\cite{white1992}. The accuracy of the approximation can be controlled systematically using the \emph{bond dimension} $M$ of the MPS, which is the size of the matrices associated with each site in the lattice; as such, the computational cost scales with the third power of the bond dimension.
The maximum bipartite entanglement that can be captured in an MPS is bounded by $\log (M)$, and therefore $M$ needs to grow exponentially with the entanglement in the system. While rigorous bounds for the scaling of MPS simulations are available~\cite{landau2014efficient,huang2015computing}, heuristically one often finds an approximately linear scaling of the computation time with system size for gapped systems.

However, if the underlying Hamiltonian is quadratic, this exponential scaling in entropy can be improved further~\cite{kraus2010,evenbly2010entanglement,schuch2012,dubail2015,haegeman2013,fishman2015,evenbly2016,haegeman2018,jahn2019holography,schuch2019}. A conventional tensor network can be understood as associating a quantum state with the vertices of a graph (which may or may not be the underlying lattice); the degrees of freedom on these states are associated with edges of the graph and can be physical or auxiliary. The physical quantum state is recovered by projecting the auxiliary degrees of freedom on each edge of the graph onto a maximally entangled state. One can now choose these quantum states associated with the vertices of the graph to be Gaussian states, i.e. states satisfying Wick's theorem, and choose the maximally entangled state that the edges are projected onto as a Gaussian state as well. In this case, the physical state being represented is Gaussian as well, and the entire computation can be performed in terms of covariance matrices of the states. This representation inherits most properties of general tensor network states; however, the exponential scaling with the entanglement entropy is replaced by a polynomial scaling, i.e. the ansatz is exponentially more efficient in terms of its scaling with entanglement entropy. This construction was used to obtain practical, efficient algorithms for one-dimensional systems of free fermions using Gaussian fermionic matrix-product states (GFMPS) in Ref.~\onlinecite{schuch2019}; these methods form the basis of the efficient gHF calculations presented in this paper.


On a technical level, a GFMPS is obtained by associated to each site $i$ on a chain a pure Gaussian state $\ket{\gamma_i}$ with covariance matrix $\gamma_i$. The fermionic modes on each state can be assigned to three groups: physical modes and auxiliary modes connecting to the left and right. These auxiliary degrees of freedom can be thought of as capturing entanglement to the left and right of the system, respectively, and the physical state is obtained by projecting the right auxiliary modes on site $i$ with the left auxiliary modes on site $i+1$ onto a maximally mixed state (often referred to as "tracing out" or "contracting"), so that only physical modes are left. The number of auxiliary modes on each bond, which we denote as $\chi$ and which should be viewed as hyperparameter refining the ansatz similar to the bond dimension $M$ for conventional MPS, bounds the bipartite entanglement in the state by $S \leq \chi \log \sqrt{2}$, i.e. the maximal entanglement is linear rather than logarithmic in the case of conventional MPS.


As a crucial ingredient for practical calculations, Ref.~\onlinecite{schuch2019} describes the canonical form for GFMPS, efficient computation of the total energy as well as a way to express the total energy as a linear function of a local tensor. These components together allow for a straightforward generalization of standard MPS techniques, such as the DMRG algorithm, which (starting from an initial guess for the state, for example a completely random state) finds an approximation of the ground state of the system by iteratively optimizing each tensor (or pairs of tensors) in the MPS. This optimization is swept back and forth across the system until convergence is reached.
While a detailed review of the technical aspects of GFMPS calculations is beyond the scope of this manuscript, we review some key aspects of the GFMPS method in Appendix~\ref{app:gfmps}.

For the discussion of our numerical results below, an important practical difference between conventional MPS and GFMPS is that in the latter case, it can be advantageous to group several physical sites together and form a lattice of such blocks. We will therefore typically refer to a block of $B$ sites, which is a single site in the GFMPS but encompasses $B$ physical sites.
Choosing $\chi$ and $B$ must be done carefully, and one must generally ascertain convergence with respect to $\chi$. For a given $\chi$, it is typically close to optimal to choose blocks of size $2\chi$ (if the goal is to minimize memory) or $\chi$ (if the goal is to minimize computation time).

\subsection{gHF using GFMPS}

\begin{figure}

\begin{algorithm}[H]
  \caption{GFMPS gHF}
\begin{algorithmic}
\Function{Gfmps-gHF}{T, U}
\State $\gfmps \gets \textrm{random initial state}$
\State \textsc{GfmpsDmrg}$(T,\gfmps)$;
\State $\Gamma = \textsc{ExtractGamma}(\gfmps)$;
\State $E_0 \gets \infty$
\For{$s \gets 1$ to $maxIter$}
  \State $F \gets 6U_{ij,kl}\Gamma_{k,l} + T_{i,j}$;
  \State \textsc{GfmpsDmrg}$(F,\gfmps)$;
  \State $E_{\textrm{new}} \gets (F_{i,j}+T_{i,j})\Gamma_{j,i}/2$
  \State $\Delta E \gets E_0 - E_{\textrm{new}}$
  \State If $|\Delta E| < \Delta E_\mathrm{target}$, \textbf{break};
  \State $\Gamma = \textsc{ExtractGamma}(\gfmps)$;
  \State $E_0 \gets E_{new}$
\EndFor
\EndFunction
\end{algorithmic}
\end{algorithm}
\caption{Pseudo-code description of the gHF iteration using a GFMPS-based solver. \label{fig:ghf-algorithm}}
\end{figure}

In the full solver, the gHF iteration forms an outer loop; its pseudocode is shown in Fig.~\ref{fig:ghf-algorithm}. In each iteration, it queries the covariance matrix $\Gamma^0$ from the underlying GFMPS representation, builds an effective potential $F$ from the covariance matrix $\Gamma^0$, and then passes this new potential to the DMRG solver to obtain an updated $\Gamma$ in GFMPS form. In the inner loop, several sweeps of the DMRG optimization are performed to obtain the lowest-energy GFMPS for a given Fock matrix $F$. The GFMPS is re-used as the initial state for the DMRG solver in the next gHF iteration in order to speed up convergence.

In principle, the entire dense covariance matrix $\Gamma$ can be extracted from a GFMPS.
However, this would require $O(N^2)$ memory and negate the time and memory savings of the GFMPS approach. We focus on a local Hamiltonian, where $T$ and $U$ connect each block to only a small number of other blocks. Then we only need to extract a block-space $\Gamma$, populating the blocks that are connected by $T$ and $U$. Strictly local models like the Hubbard model have only intra-block quartic terms ($U$), and inter-block quadratic terms ($T$), so that a sparse $\Gamma$ can be extract in $O(N)$ time; the same asymptotic scaling will be preserved for models with finite but bounded range (e.g., next-nearest-neighbor interactions). Conversely, a non-local interaction such as an unscreened Coulomb interaction with $1/r$ decay would add terms between all pairs of blocks, and thus require computing all elements of $\Gamma$ and lead to a dense $\Gamma$ and $F$. In this unfavorable case, the GFMPS-based approach would recover the computational cost of the dense approach. In some cases, for example for screened Coulomb interaction of the form $e^{-r/\xi}/r$, it may be possible to introduce a sharp cutoff and set all interaction terms beyond this distance to zero in order to recover the linear scaling.

Pseudocode for the subroutines \textsc{GfmpsDmrg} and \textsc{ExtractGamma} can be found in Appendix~\ref{app:pseudocode}. It is important to note that the same $\gfmps$ object is being used across iterations, and the previous state computed by \textsc{GfmpsDmrg} is used as input to the next \textsc{GfmpsDmrg}. After the first one or two iterations, the effective potential $F$ will not change much, so the previous state of $\gfmps$ is a good initial state for the DMRG solver.

\section{Results}

\subsection{Model}

To demonstrate the efficacy of the approach, we study the Hubbard model on a two-dimensional rectangular lattice with a quadratic anisotropic trapping potential, loosely modeling trapped quantum gases~\cite{clark2009trap,debnath2018blockade,Sterling2014}. The Hamiltonian is given by
\begin{eqnarray}
H &=& H_0 + H_{\rm int} + H_{\rm trap} \\
H_0 &=& -t\sum_{\langle \mathbf{x}, \mathbf{y}\rangle,\sigma} a_{\mathbf{x}\sigma}^\dagger\, a_{\mathbf{y}\sigma}  - \mu \sum n_{\mathbf{x}\sigma}\\
H_{\rm int} &=& U \sum_{\mathbf{x}} \left(n_{\mathbf{x}\uparrow}-\frac{1}{2}\right)\left(n_{\mathbf{x}\downarrow}-\frac{1}{2}\right) \\
H_{\rm trap} &=& \sum_{x,y,\sigma} (V_x x^2 + V_y y^2) n_{x,y,\sigma},
\end{eqnarray}
where $a_{\mathbf{x}\sigma}^\dagger$ creates a fermion of spin $\sigma$ on site $\mathbf{x}=(x,y)$ of the lattice, by $\langle \mathbf{x}, \mathbf{y} \rangle$ we denote pairs of nearest-neighbor pairs, and $n_{\mathbf{x}\sigma} = a_{\mathbf{x}\sigma}^\dagger a_{\mathbf{x}\sigma}$. Here, $t$ denotes the hopping strength, $\mu$ sets the chemical potential and thus controls the filling of our system (noting that due to the Majorana representation being used in our method, we don't fix the particle number), $U$ is the strength of the on-site Hubbard interaction, and $V_x$ and $V_y$ control the properties of the harmonic trap.  We quote all energy scales below in units of the hopping $t$.

In addition to the parameters of the physical model, there are the parameters of the method. Both dense gHF and GFMPS are run until $\Delta E < 10^{-3}$, where $\Delta E$ is the energy difference after subsequent iterations. The GFMPS has additional parameters for the bond dimension $\chi$ and block size $B$. Lengths were picked to always be multiples of $B$, so that all blocks were equal size. We generally performed 4 GFMPS DMRG sweeps per gHF iteration and use the single-site DMRG algorithm~\cite{white2005}.

\begin{figure}
\begin{center}
\includegraphics[width=\columnwidth]{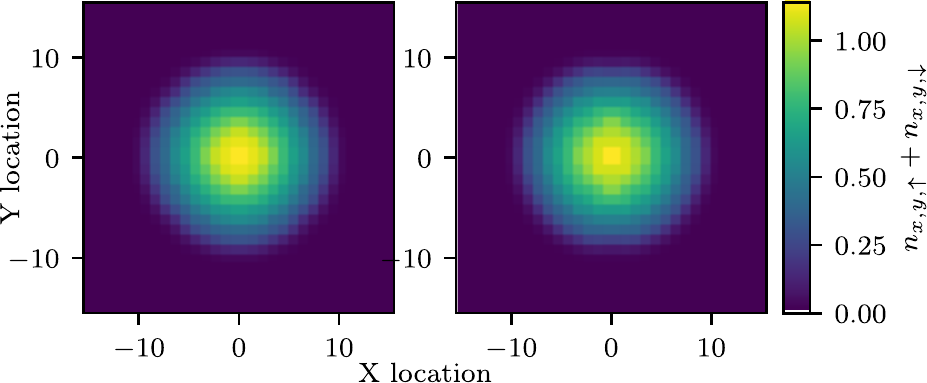}
\includegraphics[width=2.2in]{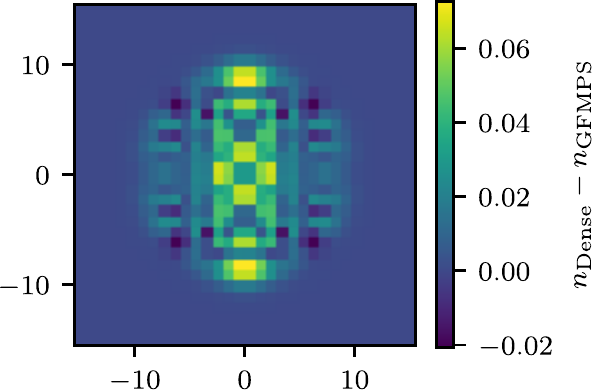}
\end{center}
\caption{Top-left: filling fraction over space with standard Hartree-Fock. Top-right: with GFMPS accelerated method. Bottom: difference between top two, contrast enhanced 13x. \label{fig:comp_3232} }
\end{figure}


\begin{figure}
    \centering
    \includegraphics[width=\columnwidth]{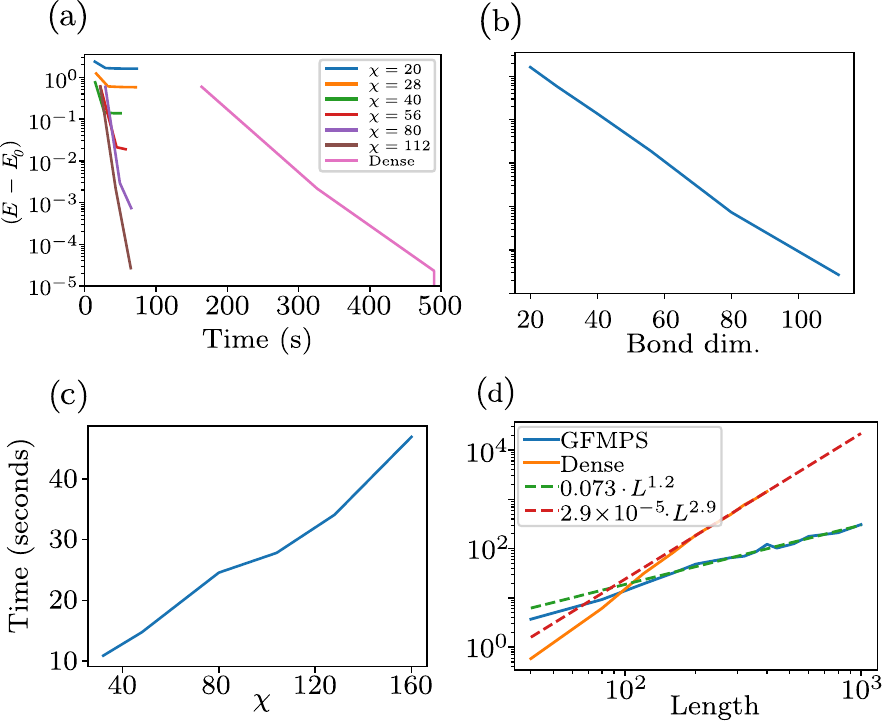}
    \caption{
    (a) Convergence of the energy as function of CPU time for various bond dimensions $\chi$ as well as for the dense Hartree-Fock solver. Here we have used $L=280, U=0.4t$. A bond dimension of $\chi=80$ sufficed to achieve similar energy to the dense solver ($<10^{-3}$ error), but ran 7.5x faster. Energy is relative to the final result of the dense computation, which converged on the 4th iteration after 653~s. (b) Convergence of the energy estimate as function of bond dimension. (c) Comparing the time for one Hartree-Fock iteration at different bond dimensions, for $L=400t, U=0.4t$. (d) Time required to run Hartree-Fock to convergence ($\delta < 0.001$) on varying system lengths. The standard dense approach displays roughly $O(n^3)$ time, while the GFMPS scales close to linearly. Dashed lines are the lines of best fit (power law fits).
    }
    \label{fig:convergence_vs_time}
    \label{fig:time_growth}
\end{figure}

\subsection{Square systems}

While the GFMPS approach is much better suited to quasi-one-dimensional systems, i.e. where the length $L$ far exceeds the width $W$, we first test the accuracy of the approach for a square system with $W=L = 32$. The parameters of the Hamiltonian were chosen as $U=0.4t$, $V_x=V_y=0.02t$, $\mu=0.3t$, $\chi=32$, $B=8$. This puts it in the weakly repulsive regime $0 < U/t < 1$. The GFMPS DRMG method found a state of energy -6791.37 and peak filling 1.140, while the full dense method found -6793.47 and peak filling 1.169. The largest difference in filling was just off-center, with 0.072. This gave agreement within a relative error of $10^{-4}$ for the energy and about $6 \cdot 10^{-2}$ for the filling. Shown in Fig.~\ref{fig:comp_3232} are the densities of GFMPS and dense solution in the top two panels, and the difference between the two in the bottom panel. We can see that the 90-degree rotational symmetry of the physical system is broken by choosing how the physical system is mapped onto the one-dimensional arrangement of the GFMPS. In Fig.~\ref{fig:comp_3232}, the sites of the GFMPS are arranged along the horizontal direction of the system. The vertical and horizontal reflection symmetry still remains in the GFMPS solution.

\subsection{Computational performance for quasi-one-dimensional systems}

To evaluate the performance benefit of the GFMPS approach for quasi-one-dimensional systems, we turn our attention to systems of fixed width $W=4$ and varying length $L$. For the other parameters, we choose $U=0.4t$, $V_x=V_y=(6/L^2)\,t$, $\mu=0.3t$, $\chi=4$, $B=8$. Each run of the DMRG used 4 sweeps. The potential was chosen to scale $V_x \sim L^{-2}$ so that the fraction of the trap occupied by particles stays roughly constant, with the potential rising from 0 in the center to 1.5 at the edges.

The first test was to see how our method compares to dense generalized Hartree-Fock. Because it is a variational approach, the minimum energy attained is our primary figure of merit. We also want to ensure that we see the linear scaling of computation time with the length of the trap for the GFMPS-based calculations, as compared to a cubic scaling for conventional, dense Hartree-Fock.

First, we held $L=280$ fixed and observed accuracy and runtime with different bond dimensions. We found that at $\chi=80$, the error in energy was $< 0.001$, which represents capturing almost all the energy that Hartree-Fock can. The GFMPS computation took only 65 seconds, as opposed to 490 seconds with dense Hartree-Fock.
This represents a 7.5x speedup. Results with other bond dimensions are shown in panels (a)-(c) of Fig.~\ref{fig:convergence_vs_time}.

The same value $\chi=80$ was then used across a broad span of lengths to see how the computation time scaled with system size for otherwise fixed parameters. Our results are shown in Fig.~\ref{fig:convergence_vs_time}(d). We expect that the GFMPS scales approximately linearly with length, while the dense method, which requires diagonalizing an $O(L)$ size matrix, would scale cubically. Fitting power laws $t = a\cdot L^p$ to each yielded exponents of $1.20$ for the GFMPS and $2.95$ for the dense methods, in good agreement with expectations. Despite holding $\chi$ fixed, the error in energy did not increase significantly, staying below $10^{-3}$.

\subsection{Repulsive case}

\begin{figure}
\begin{center}
\includegraphics[width=\columnwidth]{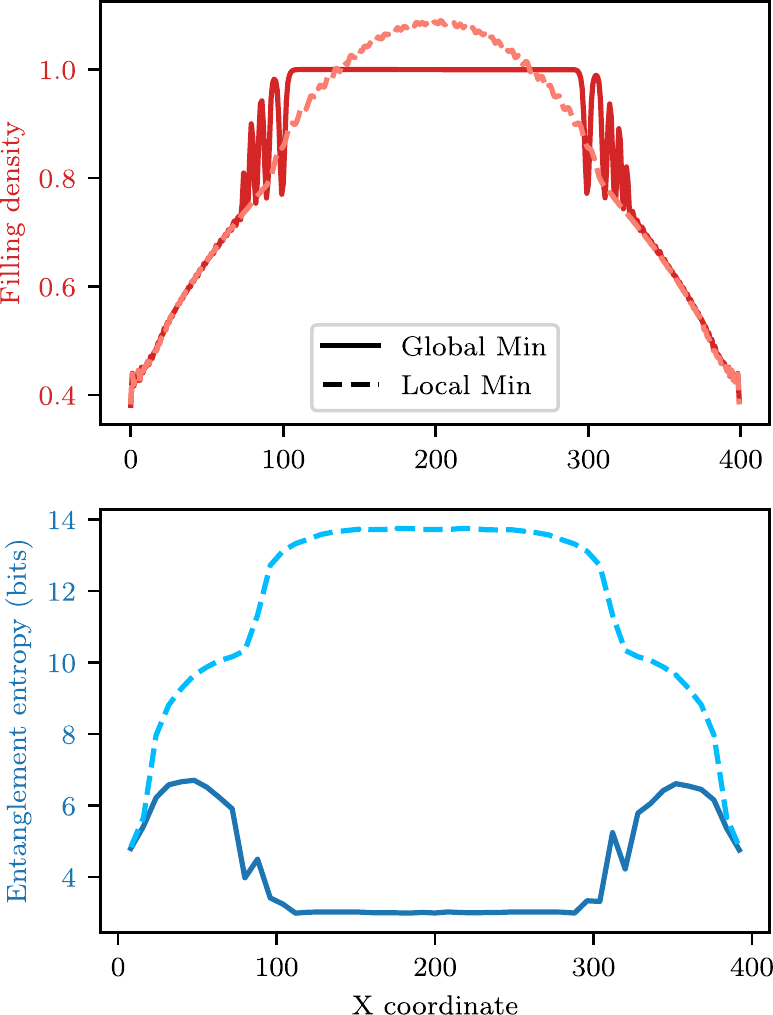}
\end{center}
\caption{Comparison of the metastable state found by standard Hartree-Fock iteration, and the true global minimum HF configuration. Red lines: Filling fraction along the length of the system. Blue lines: entanglement entropy across different cuts of the system. Solid line is the local minimum which fails to avoid the repulsive energy penalty, and has accordingly higher entanglement entropy in middle of the system. \label{fig:local_v_global} }
\end{figure}

\begin{figure}
\begin{center}
\includegraphics[width=\columnwidth]{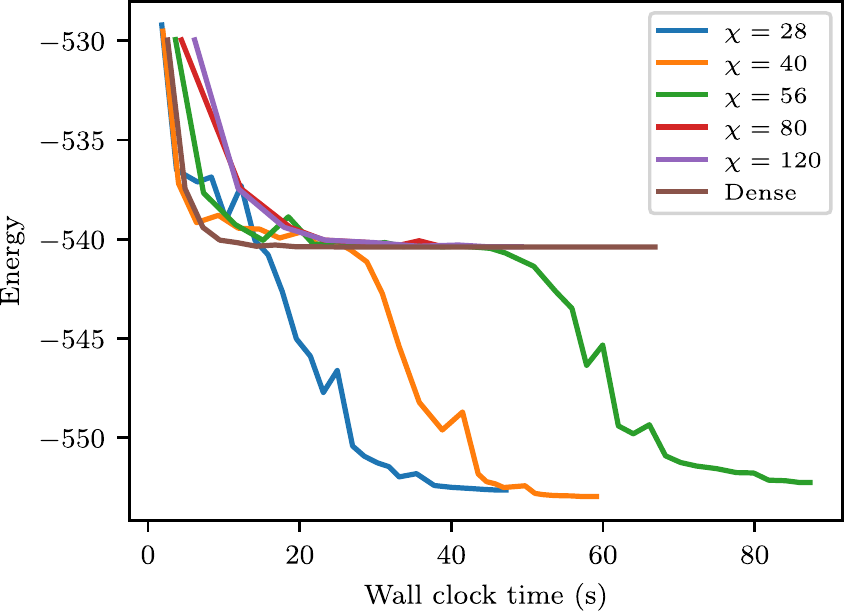}
\end{center}
\caption{Evolution of energy over time as GFMPS DMRG is run with different bond dimensions. It was expected that smaller bond dimensions would fall more quickly, but bottom out at higher energy. We found instead that higher bond dimensions became stuck at a higher energy, due to a local energy minimum.\label{fig:chi_tunneling} }
\end{figure}


Having established the improved performance of the GFMPS-based solver for weakly interacting systems, we now investigate whether the expected behavior is found also in more strongly interacting cases, starting with the case of repulsive interactions. To this end, we increase the interaction strength to $U/t = 3.0$.
The phase diagram of the translation-invariant Hubbard model is well-understood~\cite{LiebWuNoTransition,essler2005} and it is known that as $\mu$ varies there are separate partially-filled (compressible) and half-filled (incompressible) phases. The half-filled phase occurs in a region of chemical potential $\mu \in [-\mu_0, \mu_0]$ centered around the half-filled point $\mu=0$. The critical value $\mu_0$ is given by~\cite{essler2005,LiebWuNoTransition}
\begin{align}
\mu_0 &= -2 + \frac{u}{2} + 2 \int_0^\infty \frac{d\omega}{\omega}\frac{J_1(\omega)e^{-\omega u/4}}{\cosh(\omega u/4)}\\
 &\approx \exp(-6/u)
\end{align}
with $u=U/t$ and the approximation good for $u \ll 1$.
By creating an effective $\mu > -\mu_0$ in the center of our trap and $\mu < -\mu_0$ on the edges, we should see distinct regions appear in the same system.

We ran with system parameters of $U=3.0t$, $V_x=V_y=6/L^2\,t$, $\mu=0.3t$, $\chi=40$, $B=8$. The GFMPS produced a solution in line with the two-phase result we expected. Its density is shown as the solid line in the top panel of Fig.~\ref{fig:local_v_global}. As anticipated, we find an expected region of unit filling in the center of the trap and a continuous decay to zero filling towards the edge.

Comparing the result of the GFMPS-based calculation with the result of the dense solver, we find that the latter converged to a state with considerably higher energy. Furthermore, as shown in the dashed line in the top panel of Fig.~\ref{fig:local_v_global}, the state did not exhibit the extended plateau of unit filling in the center of the system. This was surprising as we generally view the GFMPS as a more restricted ansatz, and therefore expect the dense solver to produce lower energies. In this case, however, it turns out that the dense solver becomes trapped in a stable yet unphysical fixed point of the Hartree-Fock iteration, a local minimum which is avoided by the GFMPS-based solver.

To understand why the GFMPS is able to avoid this fixed point, we compared the entanglement entropy of the global minimum and the local minimum, as shown in the bottom panel of Fig.~\ref{fig:local_v_global}. The entanglement of the local minimum is much larger than that of the global minimum, which in the center of the trap potential corresponds to an incompressible and thus weakly entangled state. This hints at why the GFMPS is able to avoid this local minimum: like all tensor-network based approaches, it is (at finite bond dimension) biased towards low-entanglement solutions,
thus making it more likely to find the incompressible state.


To confirm this, we reran the GFMPS solver with much larger bond dimensions, where its behavior should more closely resemble that of the dense solver. Results are shown in Fig.~\ref{fig:chi_tunneling}. We find that indeed for bond dimensions $\chi \geq 80$, the GFMPS-based solver becomes trapped in the same local minimum as the dense solver.
Intermediate bond dimensions may become trapped in this local minimum for a few sweeps, but eventually find the global minimum. Overall, this suggests that it may in some situations be beneficial to limit the bond dimension of the GFMPS at least in initial sweeps of the self-consistent iteration.

\subsection{Attractive case}
When $U < 0$, the interaction between fermions is attractive, and we expect the appearance of finite superconducting pairing as measured by $a_{i,\uparrow}a_{i,\downarrow} + a^\dagger_{i,\uparrow}a^\dagger_{i,\downarrow}$. At the half-filled point $\mu=0$, the Hubbard model on bipartite lattices has a $U \to -U$ symmetry corresponding to applying $(a_{i,\uparrow} + a^\dagger_{i,\uparrow})(a_{i,\downarrow} + a^\dagger_{i,\downarrow})$ at every other site. To study specifically the behavior of the superconducting phase, we thus study densities away from half-filling.

We simulate at $\mu=0.5$, $L=2000$, varying $U/t$ in order to see both the weakly- and strongly-interacting cases.
\par As expected, we observe that the convergence is fast for weak interactions like $U=-0.5t$, where the Hartree-Fock procedure does not modify the potential as greatly between iterations. $U=-4t$ converged more slowly in the middle, as discussed in the previous section, likely for similar reasons of gradually adjusting the potential. All solutions showed density oscillations, especially pronounced in the vicinity of the $\braket{n_i} = 1$ point (the $x$ location with a filling of approximately one fermion per site). Running the Hartree-Fock iteration for many more steps gradually reduced the amplitude of the oscillations, but they did not go away, suggesting that these Friedel-like oscillations are genuine physical effects, but that the search procedure may be prone to overestimating them.

It is well-known that many qualitative features of the superconducting phase are well-captured by the BCS~\cite{bardeen1957theory} mean-field solution, which can be viewed as a more restricted version of the gHF ansatz. For the translationally-invariant case, the mean-field solution can be obtained semi-analytically by solving the gap equation, which relates the filling fraction $n$, superconducting gap $\Delta$ and the interaction strength $U$, and in one dimension takes the form
\begin{eqnarray}
    \xi(k) &=& -2\cos(k) - \mu - U\left(n - \frac 12\right) \\
    n &=& \frac{1}{2\pi} \int_{0}^{2\pi} \Theta\left(-\xi(k)\right)\, dk \\
    \frac{2}{|U|} &=& \int_{0}^{k_F} \frac{dk}{\sqrt{\xi(k)^2 + \Delta^2}}.
\end{eqnarray}

To compare to our numerical gHF solution, we can consider what we refer to as "local gap approximation" (following the "local density approximation" widely used in density-functional theory calculations), where we approximate the solution at each point in space by the solution of the (translationally-invariant) gap equation for the parameters at that point in space. This should be appropriate in the limit where the potential varies very slowly compared to the coherence length of the superconductor.


\begin{figure}
\begin{center}
\includegraphics[width=\columnwidth]{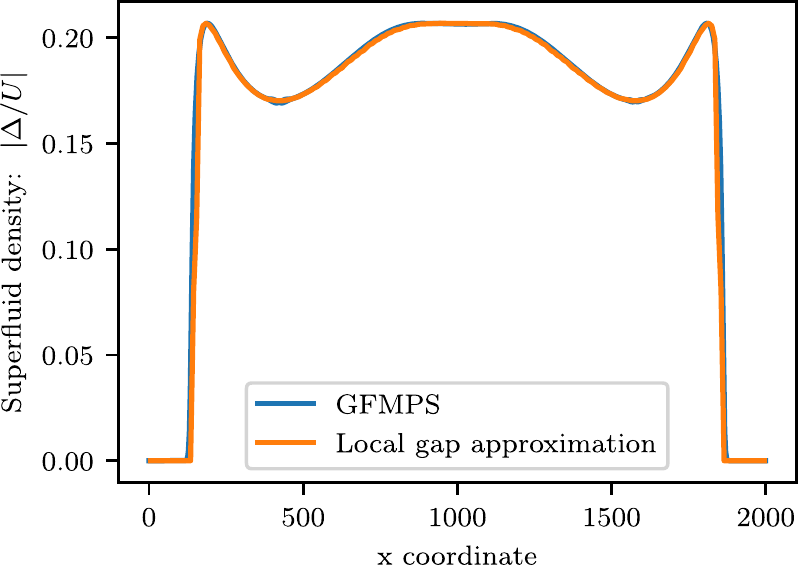}
\includegraphics[width=\columnwidth]{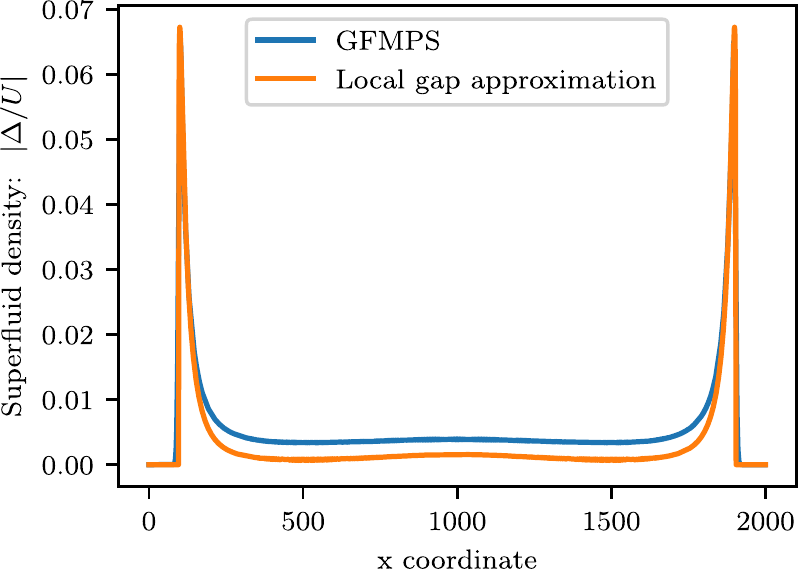}
\end{center}
\caption{Comparison of gHF-GFMPS calculation of superfluid density with predictions of superfluid density from the BCS gap equation. Both have $V=6t/L^2$. Top figure: $\mu=-1$, $U=-2$. Bottom figure: $\mu=-0.75$, $U=-0.65$. As the local gap approximation becomes more accurate as $L$ increases, a larger $L$ of 2000 was chosen for the bottom figure to show how the differences persist. \label{fig:sf_vs_invariant} }
\end{figure}

At $U=-2$, where the coherence length is on the order of a few lattice sites, this local gap approximation accurately reproduces the numerical gHF solution, as shown in the top panel of Fig.~\ref{fig:sf_vs_invariant}. The deviation is most pronounced at the edges of the system, where the local gap approximation has the superfluid density drop sharply to zero, while in the true inhomogeneous problem it tapers off over a few sites.
At the much weaker interaction strength of $U=-0.65$, which is much more representative of real-world conditions, where superconducting pairing is a weak effect compared to the Fermi energy, a very distinct picture emerges. Our results for this regime are shown in the bottom panel of Fig.~\ref{fig:sf_vs_invariant}. We find that the local gap approximation is much less accurate at capturing the inhomogeneous solution, underestimating the strength of superconducting pairing by approximately a factor of 2 in the center of the system. Our GFMPS-based approach is still able to solve this inhomogeneous system of several thousands of degrees of freedom in 113 seconds.

\section{Outlook}

In this paper, we have demonstrated that using numerical methods based on Gaussian tensor networks can accelerate computational methods that map many-body electron problems onto effectively non-interacting problems. This family includes Hartree-Fock and self-consistent BCS, which can be unified into the generalized Hartree-Fock method used here, but also other widely used approaches such as density functional theory. We thus expect this general approach to be applicable to a wide array of problems.

Application areas where an inhomogeneous real-space solution may be particularly important include systems with very large unit cells, which is a feature typically found in Moir\'e materials such as twisted bilayer graphene (tBLG)~\cite{andrei2020graphene}. Our approach seems suitable, for example, to extend recent studies of tBLG using hybrid Wannier orbitals to larger systems~\cite{hejazi2021hybrid}. Similarly, in mesoscopic device physics, inhomogeneities in the system often play an important role, and the methods put forward here suggest a pathway to realistic simulations of such structures.

\acknowledgements

We'd like to thank Kasra Hejazi for useful discussions.

\appendix

\section{Majorana form of Hamiltonians}
\label{appendixMaj}
In standard creation and annihilation operators, a quartic Hamiltonian is written
\begin{equation}
    H = \sum t_{ij}a_i a^\dagger_j + \sum u_{ijk\ell}a_ia_ja^\dagger_k a^\dagger_\ell + h.c.
\end{equation}
This can be expanded in terms of Majorana operators,
\begin{eqnarray}
     a_i \to \frac{1}{2}(c_{2i-1}+ic_{2i})\\
     a_i^\dagger \to \frac{1}{2}(c_{2i-1}-ic_{2i})
\end{eqnarray}
Then any term from $t$ becomes
\begin{equation}
\label{tTerm}
t_{ij}a_i a^\dagger_j = \frac{1}{4}t_{ij}(c_{2i-1}c_{2j-1}+ic_{2i}c_{2j-1}-ic_{2i-1}c_{2j}+c_{2i}c_{2j})
\end{equation}
Its Hermitian conjugate is
\begin{eqnarray}
t_{ij}^* a_j a^\dagger_i = \frac{1}{4}t_{ij}^*(c_{2j-1}c_{2i-1}+ic_{2j}c_{2i-1}-ic_{2j-1}c_{2i}+c_{2j}c_{2i})\nonumber\\
= \frac{1}{4}t_{ij}^*(-c_{2i-1}c_{2j-1}+ic_{2i}c_{2j-1}-ic_{2i-1}c_{2j}-c_{2i}c_{2j}+2\delta_{i,j}\nonumber
\end{eqnarray}
where $\{c_i,c_j\}=\delta_{ij}$ was used. The $\delta_{ij}$'s that arise can be discarded, as they only add some constant shift to the Hamiltonian. Adding \eqref{tTerm} to its Hermitian conjugate yields
\begin{equation}
  \frac{t_{ij}+t_{ij}^*}{4}(ic_{2i}c_{2j-1}-ic_{2i-1}c_{2j}) + \frac{t_{ij}-t_{ij}^*}{4}(c_{2i-1}c_{2j-1}+c_{2i}c_{2j})
\end{equation}
\begin{equation}
=
  \frac{i\Re[t_{ij}]}{2}(c_{2i}c_{2j-1}-c_{2i-1}c_{2j}) + \frac{i\Im[t_{ij}]}{2}(c_{2i-1}c_{2j-1}+c_{2i}c_{2j})
\end{equation}
This shows that the resulting expansion has only imaginary coefficients on the quadratic terms. The anticommutation of the Majorana operators lets us rewrite any $c_ic_j \to \frac{1}{2}\delta_{ij} + \frac{1}{2}(c_ic_j - c_jc_i)$. And then, again, we can neglect the $\delta_{ij}$ terms that lead to a constant offset in the Hamiltonian. In this way, the quadratic terms $t_{ij}a_ia_j^\dagger$ can always be transformed into
\begin{equation}
    C + \sum_{ij} iT_{ij}c_ic_j
\end{equation}
with $T$ antisymmetric and real, and some constant shift $C$.
If we apply the same substitution and expansion to the quartic terms
$$\sum_{ijk\ell}u_{ijk\ell}a_ia_ja_k^\dagger a_\ell^\dagger,$$
we will obtain additional constant terms (from e.g. $c_1c_2c_1c_2=-1$), quadratic terms ($c_1c_2c_3c_2=-c_1c_3$), and new quartic terms ($c_ic_jc_kc_\ell$, all indices distinct). When this is combined with its Hermitian conjugate, the terms combine and antisymmetrize as before, and the constant and quadratic terms are again of the form
$$C + \sum_{ij}iT'_{ij}c_ic_j$$
which can be absorbed into our other earlier quadratic term. The quartic terms remain where all indices are distinct, and the sum with the Hermitian conjugate is
\begin{eqnarray}
U'_{ijk\ell}c_ic_jc_kc_\ell + U'^*_{ijk\ell}c_\ell c_kc_jc_i\\
= U'_{ijk\ell}c_ic_jc_kc_\ell + (-1)^6 U'^*_{ijk\ell}c_ic_jc_kc_\ell\\
= (U'_{ijk\ell} + U'^*_{ijk\ell})c_ic_jc_kc_\ell = U_{ijk\ell}c_ic_jc_kc_\ell
\end{eqnarray}
The $(-1)^6$ arises from the 6 swaps necessary to reorder the $c$'s, and the resulting $U$ is completely real. Because all four $c$'s anticommute, this $U$ can be taken completely antisymmetric.

In general, a term with $k$-fermion interactions can be written with a totally antisymmetric rank-$k$ tensor. Since it will introduce $k$-choose-2 swaps when taking the Hermitian conjugate, it will be real when $k$ is a multiple of 4, and completely imaginary otherwise.

\section{GFMPS techniques}
\label{app:gfmps}

The covariance matrix $\gamma$ on the $i$'th site of a GFMPS is written as a block matrix
\begin{equation}
\gamma = \begin{bmatrix}
    \gamma_{pp} & \gamma_{p\ell} & \gamma_{pr} \\
    \gamma_{\ell p} & \gamma_{\ell\ell} & \gamma_{\ell r} \\
    \gamma_{rp} & \gamma_{r\ell} & \gamma_{rr} \\
\end{bmatrix},
\end{equation}
where the subscript labels $p$, $\ell$, $r$ denote the physical, left auxiliary, and right auxiliary modes, respectively. The $\gamma_{pp}$ block describes covariance with the $p$ sites, while $\gamma_{p\ell}$ describes covariance between the $p$ and $\ell$ sites.
%
The contraction between two such tensors is most easily illustrated for two states where the modes are arranged in two groups each; the generalization to three or more groups is straightforward. Consider two covariance matrices given, in block-form, by
\begin{equation}
    G = \begin{bmatrix}
G_{aa} & G_{ac}\\
-G_{ca}^T & G_{cc}\end{bmatrix},\quad
H = \begin{bmatrix}
H_{bb} & H_{bc'}\\
-H_{c'b}^T & H_{c'c'}\end{bmatrix}
\end{equation}
with a common subsystem $c = c'$. They can be contracted into the $ab$ covariance matrix
\begin{equation}
\begin{aligned}
G \triangleright H = &\begin{bmatrix}
G_{aa} & 0\\
0 & H_{bb}\end{bmatrix} + \\
&\begin{bmatrix}
G_{ac} & 0\\
0 & H_{bc'}\end{bmatrix}\begin{bmatrix}
G_{cc} & \mathbf{1}\\
-\mathbf{1} & H_{c'c}'\end{bmatrix}^{-1}\begin{bmatrix}
G_{ac} & 0\\
0 & H_{bc}\end{bmatrix}^T
\end{aligned}
\end{equation}

The other crucial step in DMRG is the Schmidt decomposition, where a state is split into two blocks (two physical subsystems), as accurately as possible, given the limited entanglement between the two. In a standard MPS, this is achieved by a singular value decomposition $\ket{\psi} = UDV^\dagger = \sum \ket{\ell_k}\lambda_k \ket{r_k}$; the smallest $\lambda_k$ are discarded in order to meet the bond dimension limit.
The analogous operation for fermionic Gaussian states was first derived by Botero and Reznik~\cite{BOTERO200439}. It proceeds by an SVD of the submatrix $\gamma_{\ell r}$, which contains the correlations between the two halves (but not their internal correlations):
\begin{equation}
    O\gamma_{ab}Q^T = \bigoplus_k \begin{bmatrix}
\mu_k & 0\\
0 & \mu_k
\end{bmatrix}
\end{equation}
This necessarily produces paired singular values $\mu_k$ and real orthogonal matrices $O$ and $Q$. These are related to the symplectic eigenvalues $\lambda_k$ of $\gamma_{aa}$ by $\mu_k = \sqrt{1 - \lambda_k^2}$. The Schmidt decomposition of $\gamma$ is then given by
\begin{equation}
    (O\oplus Q)\gamma(O\oplus Q)^T = \bigoplus_k \begin{bmatrix}
    0&\lambda_k&\mu_k&0\\
    -\lambda_k&0&0&\mu_k\\
    -\mu_k&0&0&-\lambda_k\\
    0&-\mu_k&\lambda_k&0
\end{bmatrix}
\end{equation}
and can be split into
\begin{equation}\gamma = L_{ac|ac} \triangleright R_{bc|bc} = 
\begin{bmatrix}0&O^T\\-O&0\end{bmatrix} \triangleright \begin{bmatrix}0&Q\\-Q^T&0\end{bmatrix}
\end{equation}
The modes where $\lambda=1$ are fully decoupled modes, and can be omitted from the $c$ index to reduce the bond dimension without altering the underlying $\gamma$. Truncating the bond dimension is achieved by setting the largest several $\lambda_k$ to 1 and discarding them, keeping only the modes with smaller $\lambda_k$ (and thus higher entanglement).

The final ingredient to reconstruct most standard algorithms for MPS in the Gaussian context is a canonical form of the GFMPS. This can be constructed in an analogous fashion to standard MPS using the SVD decomposition described above; in the case of GFMPS, it turns out that the canonical form is essential in order to make the computation of the total energy as well as the local effective Hamiltonian in the DMRG iteration efficient. For technical details of this, we refer to Ref.~\onlinecite{schuch2019}. With these tools in hand, one can perform the conventional single- and two-site DMRG algorithms to find the lowest-energy GFMPS for a given quadratic fermionic Hamiltonian.

\section{Extended Pseudocode}
\label{app:pseudocode}

\begin{algorithm}[H]
  \caption{Extracting blocks of $\Gamma$}
\begin{algorithmic}
\Function{ExtractGamma}{$\gfmps$}
\State Move $\gfmps$ to canonical form at block 0;
\State $\Gamma_{result} \gets \textrm{sparseZeros}$();
\State $carriedBlocks \gets []$;
\For{$i\gets 1$ to $numBlocks$}
  \State Add the $\Gamma_{ir}$ to $carriedBlocks$;
  \State Move $\gfmps$ to canonical form at block $i$;
  \For{$\Gamma_{jr}$ in $carriedBlocks$}
      \If{block $\Gamma_{ji}$ is needed by $T$ or $U$}
       \State Use $\Gamma_{jr}$ to compute $\Gamma_{ji}$;
        \State $\Gamma_{result}[j,i] \gets \Gamma_{ji}$;
        \State Move the 2nd index of $\Gamma_{jr}$ from $i-1$ to $i$;
     \Else
        \State Remove $\Gamma_{ji}$ from $carriedBlocks$;
      \EndIf
  \EndFor
\EndFor
\State Return $\Gamma_{result}$
\EndFunction
\end{algorithmic}
\end{algorithm}

\begin{algorithm}[H]
  \caption{GFMPS DMRG}
\begin{algorithmic}
\Function{GfmpsDmrg}{$F$,$\gfmps$}
\For{$s \gets 1$ to $maxSweeps$}
  \State Initialize effective potential $H_0$ to 0;
  \For{$i\gets 1$ to $numBlocks$}
    \State Move $gfmps$ to canonical form at block $i$;
    \State Update effective potential $H_0$ using $F$ and block $i$ of $\gfmps$;
    \State Gaussian SVD to optimize block $i$;
  \EndFor
  \State If $|\Delta E| < 10^{-3}$, \textbf{break};
\EndFor
\EndFunction
\end{algorithmic}
\end{algorithm}

\bibliographystyle{apsrev4-1}
\bibliography{biblio}

\end{document}